\documentclass{article}
\usepackage{graphicx}
\begin{document}
 
\title{{\bf STARS OF EXTRAGALACTIC ORIGIN IN THE SOLAR NEIGHBORHOOD}}
\author{{\bf T.V. Borkova, V.A. Marsakov}\\
Institute of Physics, Rostov State University,\\
194, Stachki street, Rostov-on-Don, Russia, 344090\\
e-mail: borkova@ip.rsu.ru, marsakov@ip.rsu.ru}
\date{accepted \ 2004, Astronomy Letters, Vol. 30, No. 3, P.148-158 }
\maketitle

\begin {abstract}
We computed the spatial velocities and the galactic 
orbital elements using Hipparcos data for 77 nearest 
main-sequence F--G stars with published the iron, 
magnesium, and europium abundances determined from 
high dispersion spectra and with the ages estimated 
from theoretical isochrones. A comparison with the 
orbital elements of the globular clusters that are 
known was accreted by our Galaxy in the past 
reveals stars of extragalactic origin. We show that 
the relative elemental abundance ratios of r- and 
$\alpha $-elements in all the accreted stars differ 
sharply from those in the stars that are genetically 
associated with the Galaxy. According to current 
theoretical models, europium is produced mainly in 
low-mass Type\,II supernovae (SNe\,II), while 
magnesium is synthesized in larger amounts in 
high-mass SN\,II progenitors. Since all the old 
accreted stars of our sample exhibit a significant 
Eu overabundance relative to Mg, we conclude that 
the maximum masses of the SN\,II progenitors outside 
the Galaxy were much lower than those inside it are. 
On the other hand, only a small number of young 
accreted stars exhibit low negative ratios 
$[Eu/Mg] < 0$. The delay of primordial star formation 
burst and the explosions of high-mass SNe\,II in a 
relatively small part of extragalactic space can 
explain this situation. We provide evidence that 
the interstellar medium was weakly mixed at the 
early evolutionary stages of the Galaxy formed 
from a single proto-galactic cloud and that the 
maximum mass of the SN\,II progenitors increased 
in it with time simultaneously with the increase 
in mean metallicity.

{\bf Keywords:} chemical composition of stars, subsystem 
of the Galaxy, accreted stars.

\end {abstract}

%%%%%%INTRODUCTION%%%%%%%%%%
\section*{Introduction}

In recently ears, observational astronomy has provided 
compelling evidence that not all of the stars that 
currently belong to our Galaxy were formed from a single 
proto-galactic cloud. The Galaxy captured some of the 
stellar objects at different times from the nearest 
satellite galaxies. The epoch of accretion of isolated 
individual fragments and extragalactic objects probably 
begans at the earliest formation stages of the Galaxy 
and is still going on. In particular, we are currently 
observing the disruption of a dwarf spheroidal galaxy 
in Sagittarius (dSph Sgr) by tidal forces from the 
Galaxy (Ibata et al. 1994; Mateo 1996). Four globular 
clusters are confidently associated with this galaxy: 
M\,54, Arp\,2, Ter\,8, and Ter\,7. The cluster Pal\,12 is 
far from this galaxy , but, according to the accurately 
reconstructed orbits of the two stellar systems, it was 
expelled from Sgr about one and a half billion ears ago 
(Dinescu et al. 2000). The massive globular cluster M\,54 
is generally believed to be the nucleus of the system 
(Larson 1996). In addition, it is highly likely that five 
more globular clusters belong to the Sgr system: M\,53, 
Pal\,5, NGC\,4147, NGC\,5053, and NGC\,5634 (Dinescu et al. 2000; 
Palma and Majewski 2002; Bellazini and Ferraro 2003). The 
galactic orbital elements of the clusters Rup\,106, Pal\,13, 
NGC\,5466, NGC\,6934, and NGC\,7006 also suggest that they 
were captured from various satellite galaxies 
(Dinescu et al. 2000, 2001). Freeman (1993) assumed that 
even $\omega $\,Cen, the largest known globular cluster in the 
Galaxy, which is close to the Galactic centre and has a 
retrograde orbit, was the nucleus of a dwarf galaxy in the 
past. Tsuchiya et al. (2003) showed, through numerical 
simulations, that the disruption of a dwarf satellite by 
tidal forces from the Galaxy and the emergence of its 
central cluster in the Galaxy in a highly eccentric orbit 
are quite possible. All the globular clusters whose 
extragalactic origin has been established solely from 
their spatial positions and velocities exhibit redder 
horizontal branches than do most of the Galactic clusters 
with a similar metallicity. If we assume, as it was 
previously done by Borkova and Marsakov (2000), that 
all of the low-metallicity globular clusters with anomalous 
morphology of their horizontal branches are extragalactic in 
origin, then there will be a factor of $\approx $\,1.5 more such 
clusters than low-metallicity clusters in the proto-disk halo, 
i.e., those formed from a single proto-galactic cloud. Therefore, 
accreted stellar objects constitute the bulk of the Galactic halo.
    
The theory of dynamical evolution predicts the inevitable 
dissipation of clusters through the combined actions of 
two-body relaxation, tidal destruction, and collisional 
interactions with the Galactic disk and bulge (see, e.g., 
Gnedin and Ostriker 1997). Indeed, traces of the tidal 
interaction with the Galaxy in the shape of extended 
deformations (tidal tails) have been found in all the 
clusters for which high-quality optical images were 
obtained (Leon et al. 2000). The latter authors even 
established for $\omega $\,Cen that, after the last passage 
through the plane of the disk, this cluster lost slightly 
less than one percent of its mass in the form of stars. 
Thus, even in the nearest solar neighbourhood, we may 
attempt to identify stars of extragalactic origin and to 
find possible differences in the abundances of heavy 
elements between them and the stars genetically 
associated with the entire Galaxy.

All the chemical elements heavier than boron are currently 
believed to have been synthesised in stars of various masses. 
According to the scenario suggested by Tinsley (1979), the 
presently observed lowest metallicity stars were formed 
from an interstellar medium enriched with the elements 
ejected by high mass ($M > 10 M_\odot$) asymptotic giant 
branch (AGB) stars and with the elements produced during 
their subsequent explosions as Type\,II supernovae (SNe\,II). 
The characteristic explosion time of SNe\,II after their 
formation is about 30\,Myr. These events inject $\alpha$- and 
r-elements and a few iron-peak elements. However, the 
production of the bulk of the iron began about one 
billion years after a burst of star formation, when 
stars with masses of $6-10M_\odot$ that were members of close 
binaries evolved and exploded as SNe\,Ia. The onset of 
the SN Ia explosion phase roughly coincides with the 
onset of the formation of a thick-disk subsystem. Since 
the contribution of SNe\,Ia to the synthesis of iron-peak 
elements is larger than their contribution to the 
synthesis of $\alpha$-elements, the ratio $[\alpha /Fe] \approx  0.4$, 
which is characteristic of low-metallicity stars, decreases 
to zero when going from $[Fe/H]\approx  -1.0$ (the lowest 
metallicity thick-disk stars) to solar metallicity stars 
(Edvardsson et al. 1993; Fuhrmann 1998). The abundance of 
europium, an element produced in the r-process, behaves 
similarly: the value of $[Eu/Fe] \approx  0.5$ typical of 
low­metallicity stars decreases with increasing metallicity, 
starting from $[Fe/H] \approx -1$. Both processes take place 
in stars whose final evolutionary stage is a SN\,II explosion, 
but the predominant yield of elements in different processes 
depends on the stellar mass. 

Recent studies have revealed field stars that do not 
follow this scenario of enrichment with $\alpha$- and r-elements. 
Thus, in particular, Carney et al. (1997), King (1997), and 
Hanson et al. (1998) discovered low metallicity stars with an 
$[\alpha /Fe]$ ratio much smaller than its expected value. 
Likewise, Barris et al. (2000), Mashonkina (2000, 2003), and 
Mashonkina et al. (2003) found halo stars with anomalous 
abundances of r-elements. In other words, there is a 
significant spread in relative elemental abundances of 
the two processes among stars with $[Fe/H] < -1.0$. The nature 
of this spread has not yet been completely established, because 
various scenarios for the enrichment of the interstellar medium 
with chemical elements can be realised both in isolated 
proto-galactic fragments inside a single proto-galactic 
cloud and in independent satellite galaxies. Here, we make 
an attempt to solve this question by analysing a sample of 
77 nearby stars for which reliable stellar parameters 
(including the abundances of certain chemical elements, 
ages, etc.) and their galactic orbital elements have been 
determined from high quality observational data.

%%%%%%OBSERVATIONAL DATA AND STELLAR PARAMETERS%%%%%%%%%%
\section*{Observational data and stellar parameters.}

We took the initial sample from the doctoral dissertation 
by Mashonkina (2003). It includes 77 nearby main-sequence 
F--G stars (we excluded one star, because no radial 
velocity was available for it). Most of them (66 stars) 
were selected from the lists by Fuhrmann (1998, 2003). 
Eleven more stars with $[Fe/H] < -1.0$ were specially studied 
by Mashonkina to extend the list toward halo stars. Since the 
lifetime of the stars in this spectral range on the main 
sequence is several billion ears, there are also the 
oldest stars of the Galaxy among them. In all cases, it 
was used spectra with a high spectral resolution (up to 
$\lambda /\Delta \lambda \approx 60000$) and a high 
signal-to-noise ratio (up to $S/N \approx 200$). Each 
star was observed at least twice. 

All the fundamental physical parameters of the stars 
(Teff, log\,g, [Fe/H], etc.) were determined from the same 
observational data. The effective temperatures were 
estimated from Balmer line profiles with an error of 
$\epsilon T_{eff}\pm 80$\,K. The gravitation that were 
determined by analysing the profiles of strong 
magnesium lines were almost equal to those determined 
from Hipparcos trigonometric parallaxes, and the 
error was $\epsilon$\,log\,g\,$=\pm 0.1$. The sample includes 
sonly stars with log\,g\,$\geq 3.5$. The value of 
[Fe/H] was determined with an error of $\pm 0.10$\,dex. 
The accuracy of the above parameters strongly 
affects the abundance estimates for various elements. 
The uncertainty in the [Mg/Fe] ratio (the surrogate of 
$[\alpha /Fe]$) is $\pm 0.10$\,dex. Mashonkina 
determined the relative non­LTE abundances [Eu/Fe] 
(the representative of r-process elements) with an 
error of $\pm 0.10$\,dex. Bernkoff et al. (2001) estimated 
the ages of the stars from the isochrones by Van den 
Berg (1992) and Van den Berg et al. (2000) by taking 
into account the peculiar heavy element abundances in 
each star. For several thick-disk subgiants, the age 
was determined with an accuracy of about $\pm 1$\,Gyr. 
Since the uncertainty in the ages of main­sequence 
stars can reach 2 or 3\,Gyr, the individual age 
estimates for these stars should be treated with 
caution and should be used only in the statistical 
sense.

For all the sample stars, we calculated the distances 
and spatial velocity components by using the Hipparcos 
catalogue and the radial velocity catalogue 
Barbier-Brossat and Figon (2000). The galactic 
orbital elements were calculated by using a 
multi-component model of the Galaxy that consisted 
of a disk, a bulge, and an extended massive halo 
(Allen and Santillan 1991). We assumed that the 
galactocentric distance of the Sun was 8.5\,kpc, 
the Galactic rotation velocity at the solar 
galactocentric distance was 220\mbox{km\,s$^{-1}$}, 
and the velocity of the Sun with respect to the 
local standard of rest was 
($U_\odot, V_\odot, W_\odot)$= (-10, 10, 6)\mbox{km\,s$^{-1}$}. 
A list of stars with all the parameters calculated 
and used here is given in the table. 

Despite the small size of our sample, it contains 
representatives of all Galactic subsystems 
(except the bulge). Its representativeness is 
slightly violated by the fact that the low metallicity 
stars were chosen from a larger volume of space to 
compensate for their scarcity in the solar 
neighbourhood, which is attributable to the high 
velocities of these stars (particularly their W 
component). However, even a few stars with $[Fe/H]< -1.0$ 
(22 stars) can reveal some of the global patterns of 
their behaviour because of the high accuracy of the 
parameters obtained for them.

%%%%%%CRITERIA FOR SEPARATING %%%%%%%%%%
\section*{Criteria for separating thin and thick disk stars}

\begin{figure*}
\centering
\includegraphics[angle=0,width=17cm,bbllx=100pt,bblly=10pt,bburx=650pt,bbury=300pt]{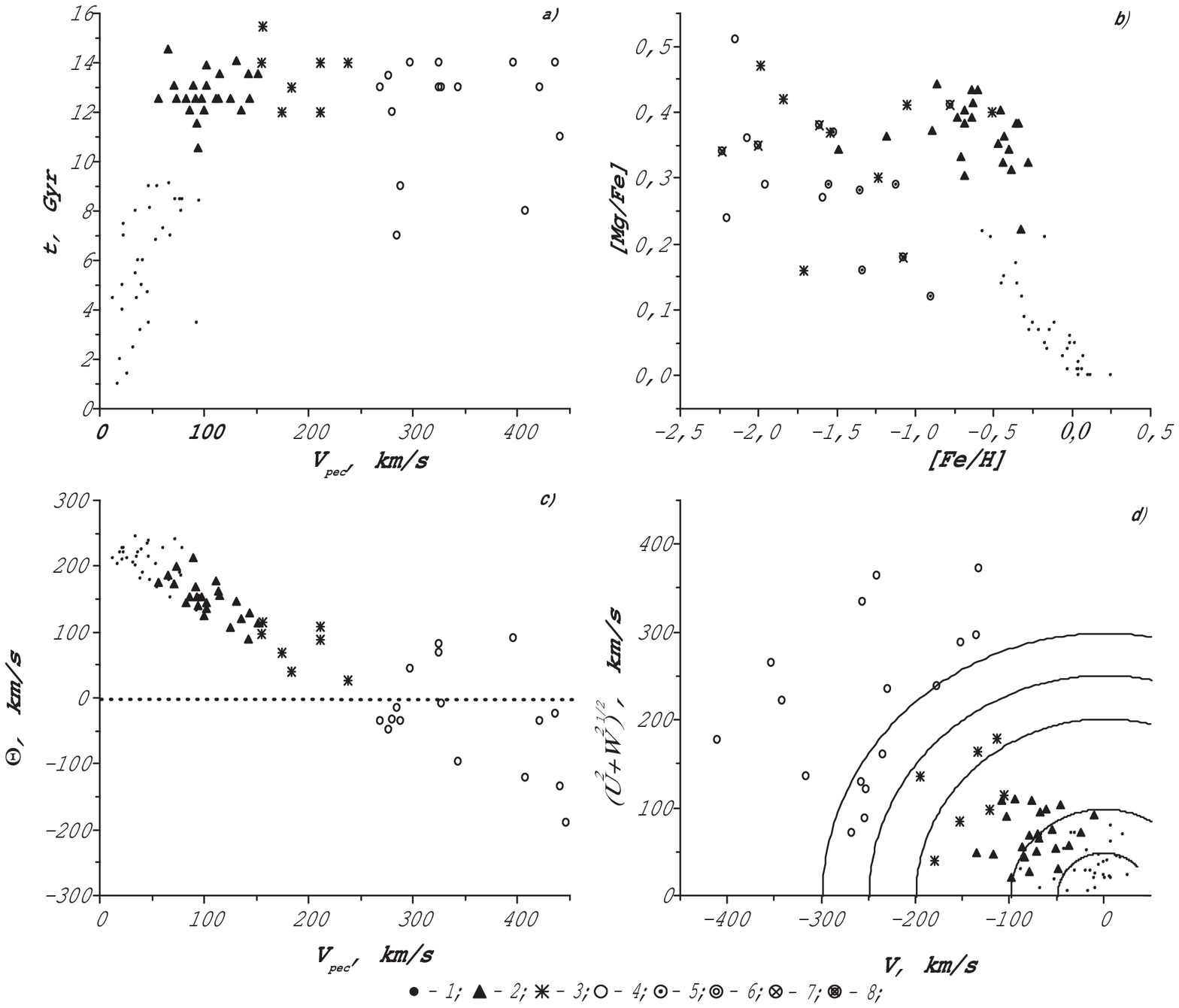}
%{\psfig{figure=fig1.ps,width=17cm,angle=-90,bbllx=332pt,bblly=42pt,bburx=61pt,bbury=761pt,clip=}}
%\includegraphics[angle=0,width=20cm,bbllx=100pt,bblly=10pt,bburx=650pt,bbury=200pt]{poyas.eps}
\caption{Correlations between the peculiar stellar 
velocities with respect to the local standard of rest 
and the stellar ages (a), between the metallicity and 
the relative Mg abundance (b), between the peculiar 
velocities and the stellar rotation velocities around 
the Galactic centre (c), and between the peculiar 
stellar velocity components: 1 -- think disk stars, 
2 -- thick disk stars, 3 -- proto-disk-halo stars, 
and 4 -- accreted halo stars. The following stars are
additionally marked in the accreted halo in panel 
(b): 5 -- stars with prograde orbits, 6 -- stars of 
the cluster from (c) with coordinates of 
$V_{pec}\approx 280$\mbox{km\,s$^{-1}$} and 
$\Theta \approx -30$\mbox{km\,s$^{-1}$}, 7 and
8---stars younger than 12.5\,Gyr
}
\label{fig1}
\end{figure*}

While analysing the properties of F--G dwarfs from his sample, 
Fuhrmann (1998) found that the Mg abundance relative to iron 
increased abruptly when going from the thin disk to the thick 
disk. This author clearly showed that there was a gap in ages 
of about 2 or 3\,Gyr between these subsystems. Marsakov and 
Suchkov (1977) first pointed out the existence of a delay 
in star formation in the Galaxy prior to the formation of 
the Galactic thin-disk subsystem. Since the thin-disk 
stars are known to have low peculiar velocities relative 
to the local standard of rest ($V_{pec}$), we separated this 
subsystem by using two criteria: $t \leq 9$\,Gyr and  
$V_{pec}\leq 100$\mbox{km\,s$^{-1}$}. A comparison of 
Figs.\ \ref{fig1}a and 1b indicate that these two criteria 
automatically separate stars with $[Mg/Fe]\leq 0.25$ (except 
one thick disk star that fell within this range). In 
the [Fe/H]--[Mg/Fe] diagram, we clearly see a gap of 
$\Delta[Mg/Fe]\approx 0.1$ between the stars of the two 
disk subsystems. However, we also simultaneously see a 
mutual overlapping of the ranges both in metallicity and 
in peculiar velocities (and, hence, in orbital sizes). 
In other words, [Fe/H] and $V_{pec}$ here are less suitable 
criteria for the individual separation of stars of these 
subsystems from one another.

Objectively, thick disk stars are much more difficult to 
separate from proto-disk halo stars genetically associated 
with them. The ages of most of the stars that do not belong 
to the thin disk lie within a narrow time interval and we can 
say nothing about the difference between the formation 
epochs of these subsystems because of the errors in its 
determination. However, based on an abundance analysis, 
Mashonkina et al. (2003) concluded that, although the time 
intervals of these subsystems overlap, star formation in 
the thick disk began about 1 Gyr later than in the halo. 
The subsystems can be separated in $[Fe/H]\approx -1.0$. 
Indeed, the metallicity distributions for globular 
clusters near this point exhibit a large deficit of 
stars (Borkova and Marsakov 2000), while the RR Lyrae 
field stars show a distinct inflection (Borkova and 
Marsakov 2002). In this case, however, a number of stars 
with circular orbits, which are characteristic of the 
disk subsystem, fall into the halo. Here, we decided to 
use the peculiar velocity as the criterion. If we 
separate the sample stars by 
$V_{pec}\approx 155$\mbox{km\,s$^{-1}$}, then only one 
star with $[Fe/H]>-1.0$ will be in the halo, and two 
low metallicity stars will be in the thick disk. Such 
stars in the thick disk are commonly called a 
low metallicity tail.

%%%%%%THE PROTODISK AND ACCRETED HALO SUBSYSTEMS%%%%%%%%%%
\section*{The protodisk and accreted halo subsystems}

There is no unique necessary and sufficient statistical 
criterion that would separate extragalactic objects. In 
each specific case, all the available parameters should 
be considered simultaneously. To stratify the field stars 
into the halo subsystems, different authors primarily 
used such parameters as the retrogradation of their orbits 
and their distances from the Galactic centre and plane. 
Naturally, the thick disk objects must first be removed 
from the sample (see above). The galactic orbital 
elements of the stars, their ages, and the abundances 
of $\alpha$-elements in them have been most commonly used as 
additional criteria (for an overview of the criteria, see 
the monograph by Carney 1999). For globular clusters, the 
morphology of their horizontal branches proved to be a good
criterion (see, e.g., Borkova and Marsakov 2000). Thus, it 
has been shown that the objects that constitute the accreted 
halo have apogalactic orbital radii larger than the
galactocentric distance of the Sun, high orbital 
eccentricities, high velocity dispersions, low-rotation 
velocities (many of them are in retrograde orbits), often 
younger ages, and an underabundance of $\alpha$-elements. 
The vertical and radial metallicity gradients are virtually 
equal to zero in the resulting subsystem. Studies of RRLyrae 
variables (Borkova and Marsakov 2002, 2003) show that a 
convenient criterion for field stars in the solar 
neighbourhood is the total peculiar velocity of the star 
relative to the local standard of rest ($V_{pec}$): when 
passing through some critical peculiar velocity (which is 
definitely higher than the circular velocity of Galactic 
rotation at the solar galactocentric distance), the 
spatial kinematics characteristics of the stars change 
abruptly. These changes suggest that all the low metallicity 
population of field RR Lyrae variables is not homogeneous, 
but consists of at least two subsystems that differ in the 
volume occupied in the Galaxy (Borkova and Marsakov 2002, 2003). 
The most compelling argument for the extragalactic origin of a 
specific star is probably a close match between its orbital 
elements and chemical composition and the analogous parameters 
of the globular cluster for which it has been firmly 
established that it was accreted by our Galaxy in the past. 

According to the hypothesis of monotonic collapse of the proto-
galaxy from the halo to the disk suggested by Eggen et al. (1962), 
the stars that are genetically associated with the Galaxy cannot 
be in retrograde orbits. On the other hand, there must also be 
stars with prograde orbits among the stars accreted by the Galaxy. 
In any case, the velocity of the accreted stars with respect to 
the local standard of rest must be very high. Therefore, the most 
natural criterion for separating accreted halo stars seems to be 
the peculiar velocity. It follows from the $V_{pec} - \Theta $ 
diagram in Fig. 1c that stars with a negative tangential velocity, 
i.e., in retrograde orbits, appear when passing through 
$V_{pec}\geq 250$\mbox{km\,s$^{-1}$}. The $V-\sqrt{U^2+W^2}$ 
diagram (see Fig.\,1d) shows how the objects of the separated 
subsystems are distributed in the plane of the peculiar velocity 
components. Within the thin and thick disks, the total peculiar 
velocity increases mainly through a decrease in the rotation 
velocities of the stars around the Galactic centre. At the same 
time, when going to the halo subsystems, the contributions 
from the other two velocity components increase sharply. Note 
the very small number of stars in the proto-disk halo compared 
to the number of stars that are assumed to have been captured 
from extragalactic space. As we pointed out above, observational 
selection is certain to have played a role here: mostly 
high­velocity stars were selected for the initial sample of 
low metallicity stars. Recall, however, that globular clusters 
also exhibit a similar ratio of the numbers, although the 
accreted halo objects were selected thereby their internal 
property, the structure of the horizontal branch, rather than 
by their spatial position (Borkova and Marsakov 2000).

Let us consider the properties of the stars that we selected 
as accreted­halo candidates in more detail. We see from the 
$V_{pec} - t$ diagram (Fig.\,1a) that there are very young stars 
among them whose ages fall even within the range characteristic 
of the thin disk. As follows from Fig.\,2 (crossed circles), the 
orbital elements are indicative of their obvious extragalactic 
origin. Four stars with prograde orbits can also raise doubts 
about their origin. In Fig.\,2, they are highlighted by circles 
with a central dot. All of them have apogalactic orbital radii 
larger than $15$\,kpc; the maximum distance of three stars from 
the Galactic plane is larger than 3\,kpc, and all stars have 
highly eccentric ($e>0.8$) orbits and old ages. These orbital 
elements fall within the range of parameters characteristic of 
the globular clusters that are assumed with a high probability 
to be accreted ones (Borkova and Marsakov 2000). Since no 
globular clusters with extremely blue branches (i.e., belonging 
to the proto-disk halo) are observed at such large 
galactocentric distances, we conclude that these four 
high velocity stars with prograde orbits may have been 
lost by the accreted clusters and be extragalactic in origin. 

The origin of yet another stellar group, the cluster of five 
stars in Fig.\,1c with coordinates $V_{pec}\sim 280$\mbox{km\,s$^{-1}$} 
and $\approx -30$\mbox{km\,s$^{-1}$} (HD\,148816,194598, 193901, 
BD\,-4$^\circ$3208, 18$^\circ$3423), can also raise doubts. Small 
circles inside large circles in Fig.\,2 highlight them. It may 
well be that at the early evolutionary stages of the proto-galaxy, 
some of the giant clouds could accidentally acquire a small 
negative rotation around the Galactic centre through their 
natural velocity dispersion. In that case, these stars must 
be oldest, their chemical composition must correspond to 
the composition of the first Galactic stars, and the maximum 
distances of the points of their orbits from the Galactic 
centre and plane must be large. Their orbital eccentricities 
turned out to be actually very high ($e \leq 0.9$). However, 
their apogalactic radii are very small, the orbits for four 
of the five stars completely lie within the solar circle, 
and their ages lie within a wide range, so three of the 
five stars are younger than $12$\,Gyr (see Fig.\,2). These 
properties are in conflict with the hypothesis that all 
these stars originated from a single proto-galactic cloud. 
Note that the orbital elements of the stars from the group 
under discussion are satisfactory agreement with those 
of the largest globular cluster $\omega$\,Cen, which, 
as was pointed out above, is probably extragalactic in origin. 

Note also the star HD\,298986, whose orbital elements are equal, 
within the error limits, to the corresponding parameters of 
the accreted globular cluster Pal\,5 (which probably belonged 
to dSph Sgr). According to the model of the Galaxy by Allen 
and Santillan (1991) used here, the parameters for the stars 
and the clusters were found to be the following. Apogalactic 
orbital radii of 22 and 19\,kpc, perigalactic radii of $2$ and 
$1.5$\,kpc, and $Z_{max}=13$ and $17$\,kpc, respectively. 
A comparison also indicates that the star and the cluster have 
not only similar $[Fe/H]$ ($-1.34$ and $-1.41$\,dex) and 
$[\alpha/Fe]$ (about $0.16$\,dex each), but also similar 
ages (about $13$\,Gyr each). 

In the next section, we show that all of the stars attributed 
to the accreted halo by their kinematics exhibit sharp chemical 
anomalies.

%%%%%%THE CHEMICAL COMPOSITION OF ACCRETED­HALO STARS%%%%%%%%%%
\section*{The chemical composition of accreted halo stars}

The $\alpha$- and r-elements are generally believed to 
be synthesised in stars with masses $M > 10 M_{\odot}$ and 
injected into the interstellar medium by SN\,II explosions. 
Therefore, the most probable [Eu/Mg] ratio for the Galactic stars 
must be equal to zero. However, the yield of $\alpha$-elements 
increases with mass of the SN\,II progenitor; the amount of 
magnesium increases by a factor of 10 to 20\,$M{\odot}$ as the 
mass of the SN progenitor changes from 13 to 25 
(Thielemann et al. 1996). On the other hand, the yield of 
r­elements is related to the explosions of the lowest mass 
type II super novae (see, e.g., Wheeler et al. 1998, 
Ishimaru et al. 2004). Mashonkina et al. (2003) performed a 
comparative analyis of the relative abundances of magnesium 
(an a-element) and europium (an r-element). They found that 
the halo stars exhibit a significant spread in Eu abundances 
relative to Mg, while the thin disk and thick disk stars 
have $[Eu/Mg]\approx 0$ with a smaller spread (the Galaxy 
was assumed to consistently of three subsystems only.) These 
authors also considered all the possible causes of this 
abundance anomaly in the halo stars and concluded that 
the bulk of the Galactic europium and magnesium were 
produced in stars of different masses, and that the 
interstellar medium was weakly mixed in the early Galaxy. 
Figure 3 shows a [Fe/H]--[Eu/Mg] diagram for the same sample, 
but the stars were stratified into four Galactic subsystems. 
We see from the figure that all the accreted halo stars 
exhibit deviations from the most probable zero [Eu/Mg] 
ratio, while among the remaining stars, only two 
proto-disk halo stars (HD\,25329 and HD\,102200) exhibit 
such deviations.

We believe that the large spread in [Mg/Fe] is an argument 
for inefficient mixing of the interstellar medium in the 
halo. However, it follows from the [Fe/H]--[Mg/Fe] diagram 
(see Fig.\,1b) that this spread is most likely associated 
with the accreted halo. Six of the seven proto-disk halo 
stars from our list have relative Mg abundances that lie 
within a narrow range, $0.32-0.47$\,dex (except the star 
HD\,122196 with an anomalously low ratio, $[Mg/Fe]=0.16$). 
Whereas the accreted halo stars occupy the range from 0.12 
to 0.51\,dex, half of them (9 of the 16 stars) have 
Mg abundances $< 0.3$\,dex. In other words, the stars formed 
far from the Galactic centre exhibit relative heavy element 
abundances that often differ from those in the stars formed 
inside the proto-galactic cloud. Clearly, not in all of the 
tidally disrupted dwarf galaxies, the star formation history 
must be the same as that in our Galaxy. Because of the large 
number of disrupted galaxies and globular clusters as well 
as the large stellar dispersion within the tidal tails from 
them, the stars that were formed from proto-stellar clouds 
with different histories of enrichment with chemical elements 
will most likely be in the solar neighbourhood. 

As was pointed out above, the accuracy of determining the 
ages for low mass main sequence stars is too low to obtain 
their statistically significant differences. Nevertheless, 
we formally divided all the accreted halo stars into two 
age groups. At the same time, the difference between the 
mean ages of these stars exceeds the error limits 
($\bigtriangleup t \approx 4 \pm 1$\,Gyr). Therefore, let 
us consider the chemical composition of the stars in each 
age group. In Figs.\,1b and 3, the open crossed circles 
highlight the accreted halo stars younger than $t < 12.5$\,Gyr. 
For approximately the same Mg abundance as that for 
protodisk-halo and thick disk stars, four of the highlighted 
young stars (except HD\,193901) exhibit an Eu underabundance 
relative to Mg. (In the star BD\,-4$^\circ$3208 
with a high ratio of $[Mg/Fe]=0.34$, the Eu abundance has 
not been determined, probably because the line of this 
element is too weak; therefore, we attributed it to the group 
with a low [Eu/Mg] ratio.) Mashonkina et al. (2003) 
excluded two stars with an Eu underabundance (HD\,34328 and 
HD\,74000)by concluding that they "did not reflect the overall 
pattern of chemical evolution of the matter in our Galaxy". 
We believe that the very young low metallicity stars with 
$[Eu/Mg] < 0$ were formed from matter that was mainly 
enriched by high mass ($M > 30M_{\odot}$) SNe\,II. According 
to existing theories for the formation of chemical 
elements, the low metallicity of these stars at a high 
$[Mg/Fe]$ ratio suggests that they are old. Thus, the ages 
estimated from evolutionary tracks are in conflict with 
their "chemical" ages. If the "isochronic" ages are actually 
accurate enough, then this contradiction can be resolved 
by assuming that the low metallicity for the anomalously 
young age of these stars is attributable to their formation 
from matter in which the primordial starburst occurred 
later than that in the single proto-galactic cloud. 
Naturally, this assumption should be further tested on 
large statistical material. The increasingly old accreted 
halo stars in Fig.\,3 exhibit a significant overabundance, 
$[Eu/Mg] > 0.2$, with the Mg abundance relative to Fe 
being lower than that observed, on average, for the 
proto-disk halo and thick disk stars: it follows from 
Fig.\,1b that six of the seven protodisk­halo stars 
have $[Mg/Fe] > 0.3$, while this ratio for eight of the 
eleven old ($t > 12.5$\,Gyr) accreted halo stars is $< 0.3$. 
The europium overabundance relative to magnesium in most 
of the old accreted halo stars suggests that the initial 
mass function of the stars formed outside the proto-galaxy 
was cut-off at high masses and began from. As a result, 
the yield of $\alpha$-elements was smaller than that within the 
single proto-galactic cloud, where the masses of the SN 
progenitors were larger by several times. This 
interpretation also accounts for the low Mg abundance 
relative to iron in old accreted stars (see Fig.\,1b). 
Indeed, the low $[\alpha/Fe]$ ratio for very old 
low metallicity stars can be more naturally explained 
by the low masses of the SN\,II progenitors than by the 
injection of iron-group elements by SNe\,Ia, because an Eu 
underabundance must then be also simultaneously observed 
in these stars. However, as we see, europium is overabundant 
in these stars. Van den Berg (2000) explained the low 
oxygen abundance in the stars of the very old 
low metallicity globular cluster M\,54, which is the centre 
of the dwarf galaxy Sgr, precisely by the deficit of 
high­mass SN\,II progenitors. It thus follows that the low 
abundance of $\alpha$-elements alone in stars cannot 
unambiguously point to slow star formation in their 
parent proto-stellar cloud, as was suggested by Gilmore and 
Wyse (1998). This quantity is often taken as a "chemical" 
indicator of a young stellar age (see, e.g., Carney et al. 
1997; King 1997; Hanson et al. 1998). Figure 1b also clearly 
shows that old stars with even lower relative Mg abundances 
appear near $[Fe/H]\approx -1.3$\,dex. Whereas nine of the 
eleven stars from the old group of accreted stars have 
$[Mg/Fe] > 0.24$\,dex, the stars HD\,298986 and 
BD\,18$^\circ$3423, being in the range 
$-1.3 \leq [Fe/H] \leq -0.9$\,dex, exhibit 
$[Mg/Fe]\approx 0.15$\,dex (i.e., there is a difference 
exceeding the error limits). Although the "isochronic" 
ages of the two stars are within the error limits, they 
are still about one billion years younger than the oldest 
stars of extragalactic origin; i.e., a time long enough for 
SN\,I explosions to occur had elapsed by the time of their 
formation. HD\,193901, which we included in the young group 
of accreted stars, lies in the same place in the diagram. 
This behaviour can be understood by assuming that the 
intergalactic matter from which all the accreted stars 
in the old group were formed acquired the primordial 
injection of heavy elements simultaneously with the 
proto-Galaxy , but from SNe\,II with masses much lower 
than those of the SNe exploded inside the proto-Galaxy 
itself. Subsequently, the star formation there was so 
slow that SNe\,Ia began to contribute appreciably to the 
iron abundance even at $[Fe/H]\approx -1.3$.

Note the two properties that the genetically associated 
stars exhibit in Fig.\,3. First, there is a spread in 
$[Eu/Mg]$ in the proto-disk halo: one of the stars that we 
attributed to this subsystem exhibits a large Eu abundance 
at a low Mg abundance (HD\,102200), while another star 
exhibits a low Eu abundance at a high Mg abundance (HD\,25329). 
Weak mixing of the interstellar medium appears to have 
actually taken place at the early evolutionary stages of our 
Galaxy, and stars were formed from clouds enriched
by ejections from SNe\,II with different masses in its 
different places. Note also that all the remaining 
genetically associated stars show a tendency for $[Eu/Mg]$ 
to decrease with increasing [Fe/H] at a relatively small 
spread (the correlation coefficient outside the 3$\sigma$ 
limits is nonzero, $r = 0.4 \pm 0.1$). This behaviour 
suggests that the maximum mass of the SN\,II progenitors 
formed inside the Galaxy increases with metallicity. The 
observed trend is also obtained if some amount of Mg is 
assumed to be additionally formed in AGB stars with, 
but none of the existing theories for the synthesis of 
heavy elements makes this assumption. Here, however, it 
should be borne in mind that a systematic bias of the Mg 
and Eu abundance estimates as a function of metallicity 
can arise, because the abundances of these elements are 
determined from lines of different ionisation stages, 
Mg\,I and Eu\,II.

%%%%%%DISCUSSION%%%%%%%%%%
\section*{Discussion}

Thus, the galactic orbital elements and the ($\alpha$- and 
r-process) elemental abundances in the stars of the 
nearest solar neighbourhood strongly suggest that 
some of them may be extragalactic in origin. The 
detected overabundance $<[Eu/Mg]> = 0.30 \pm 0.03$ in 
all the old accreted stars of our sample (see Fig.\,3) 
can be explained only by assuming that the initial 
mass function of the stars being born outside the 
Galaxy is cut-off at high masses. The simple 
assumption about weak mixing of the intergalactic 
matter with a single initial mass function for the 
entire local system seems to be less tenable. Indeed, 
the natural isolation of the explosion sites of SNe\,II 
with different masses from one another must give rise 
to the next generations of stars, with an overabundance 
of both europium and magnesium. Moreover, since they 
yield of $\alpha$-elements becomes well ahead of the
yield of r-elements as the mass of the SN\,II progenitor 
increases, we must detect Mg-over-abundant stars with a 
higher probability than Eu-over-abundant stars. However, 
interference from observational selection is possible 
here: the Eu lines used to estimate the Eu abundance 
in a star will be so weak that they will be lost in 
spectral noise. Thus, the Eu abundance cannot be determined 
in a star where the amount of this element is small. 
As a result, a deficit of stars with $[Eu/Mg] < 0$ can 
arise in the sample. In our sample, we attributed 16 
stars to the accreted halo. Eight and three of these 
stars exhibit $[Eu/Mg] > 0.2$ and less than zero, 
respectively. For five stars, the Eu abundance has not 
been determined. The metallicity of all five stars is 
approximately two orders of magnitude lower than its 
solar value. For two of them, $[Mg/Fe] < 0.3$\,dex, 
while for the other three stars, the Mg abundance is 
comparable to its abundance in the protodisk-halo and 
thick-disk stars. We may assume a relative Eu 
underabundance in the last three stars and an Eu 
overabundance in the first two stars. Thus, we have 
ten stars with an Eu underabundance and six stars 
with an Eu overabundance relative to Mg; i.e., only 
about a third of the stars could be formed from matter 
enriched by the explosions of high­mass SNe\,II. Four 
of the above six stars with $[Eu/Mg] < 0$ are younger 
than 12.5\,Gyr (BD\,-4$^\circ$3208, HD\,74000, 
HD\,34328, and HD\,148816); above, we have assumed a 
later initial burst of star formation in the part of the 
extragalactic interstellar medium from which they 
were formed. Hence, high-mass SN progenitors began to 
explode in extragalactic space much later. The small 
number of accreted stars with an Eu underabundance in 
our sample most likely implies that high mass SN 
progenitors outside the Galaxy do not determine the 
situation. However, for a Salpeter mass distribution of 
stars, SN  progenitors with $M > 30 M_{\odot}$ (they are 
believed to be the main suppliers of magnesium) 
must contaminate a much larger volume of the interstellar 
medium with $\alpha$-elements than the volume that SNe\,II with 
masses of $\approx 10M_{\odot}$ (the main suppliers of 
europium) contaminate with r-elements. This is because 
the yield of $\alpha­$elements in high-mass SN progenitors 
is a factor of about 20 larger than that in low mass 
SN progenitors (Thielemann et al. 1977), while 
they yield of r-elements decreases (Wheeler et al. 1998). 
Thus, we believe that weak mixing of the extragalactic 
medium can explain only the general spread in $[Eu/Mg]$ 
ratios in the accreted halo, while the dominance of 
stars with an Eu overabundance relative to Mg in it 
is probably attributable to the lower masses of the 
SN\,II progenitors outside Galaxy than those in the 
Galaxy.

{\bf Acknowledgements}: 
%\begin{acknowledgements}
We wish to thank L.I. Mashonkina for the opportunity 
to use the results of her doctoral dissertation, for 
helpful consultations and valuable remarks.
%\end{acknowledgements}

\newpage

\begin{figure*}
\centering
\includegraphics[angle=0,width=17cm,bbllx=100pt,bblly=10pt,bburx=750pt,bbury=650pt]{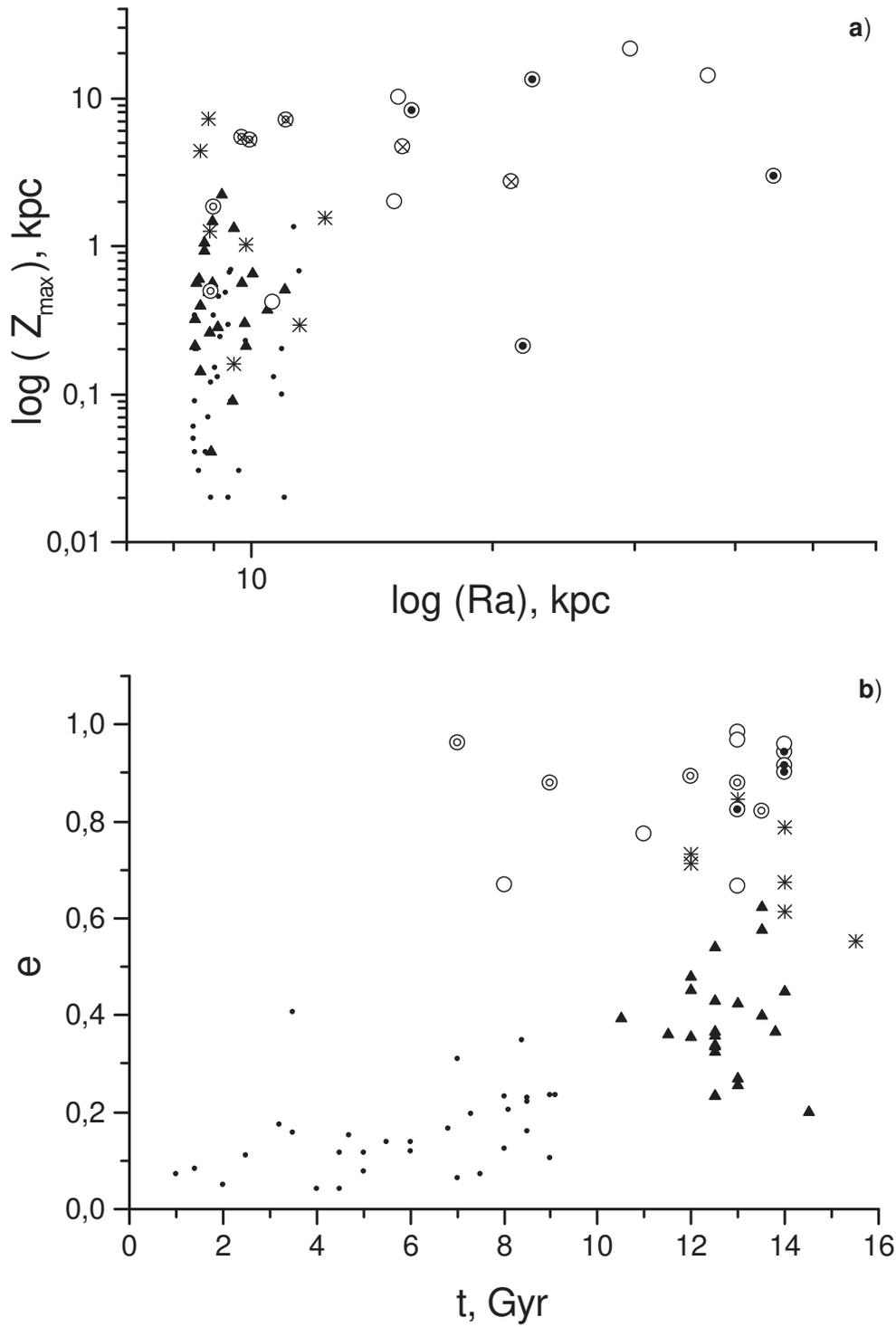}

%\centerline{\psfig{figure=fig2.ps,width=17cm,angle=-90,bbllx=389pt,bblly=42pt,bburx=60pt,bbury=763pt,clip=}}
\caption{Correlations between the maximum distances of the 
points of stellar orbits from the Galactic centre and plane 
(a) and between the ages and the orbital eccentricities of 
stars (n). The notation is the same as that in Fig.\,1.}
\label{fig2}
\end{figure*}

\begin{figure*}
\centering
\includegraphics[angle=0,width=17cm,bbllx=100pt,bblly=10pt,bburx=750pt,bbury=650pt]{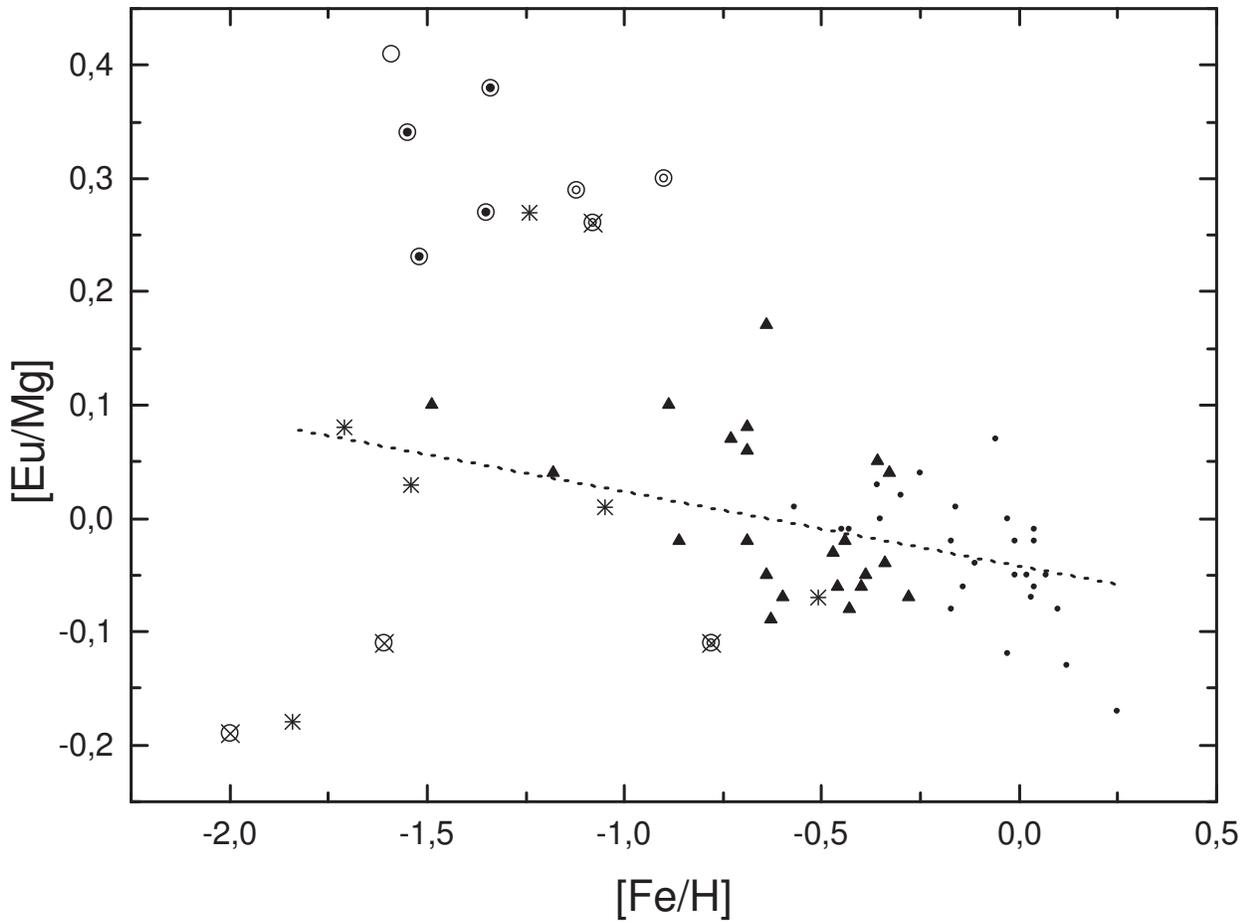}

%\centerline{\psfig{figure=fig2.ps,width=17cm,angle=-90,bbllx=389pt,bblly=42pt,bburx=60pt,bbury=763pt,clip=}}
\caption{Correlation between the iron abundances and the 
$[Eu/Mg]$ ratios for stars in the solar neighbourhood. The 
notation is the same as that in Fig.\,1. The dotted line 
represents an rms regression for the genetically 
associated stars ($r =0.4 \pm 0.1$).}
\label{fig3}
\end{figure*}

\begin{table*}
\caption[]{\bf{Chemical composition and galactic orbital elements of the
nearest field stars}}
\tabcolsep0.9mm

\begin{tabular}{ccccccccccccc}
\hline
HD/BD & [Fe/H] & [Mg/Fe] & [Eu/Fe]& t, Gyr & U,  & V, & W, & $V_{pec}$, &
Ra, & $Z_{max}$, & e & code \\
      &  dex   &  dex    &  dex   &      & km/s&km/s&km/s& km/s       & 
kpc & kpc      &   &     \\
\hline

    400 & -0.25 & 0.08 & 0.12 &  6.0 & -27.2 &  -9.2 & -8.4 &  37.2 &  9.7 & 0.0 & 0.12 & 1 \\
   3795 & -0.64 & 0.39 & 0.56 & 13.8 &  50.4 & -88.9 & 45.6 & 102.6 &  8.8 & 0.9 & 0.37 & 2 \\
   4614 & -0.30 & 0.09 & 0.11 &  7.0 &  29.9 & -10.0 &-16.8 &  22.6 &  9.1 & 0.1 & 0.06 & 1 \\
   9407 &  0.03 & 0.01 &-0.06 &  7.3 & -50.4 &  -2.0 &  1.4 &  61.3 & 10.9 & 0.1 & 0.20 & 1 \\
  10519 & -0.64 & 0.43 & 0.38 & 13.5 &  96.5 & -77.6 & 29.3 & 115.3 &  9.8 & 0.6 & 0.40 & 2 \\
  10697 &  0.10 & 0.00 &-0.08 &  6.8 & -36.0 & -27.5 & 16.2 &  54.0 &  9.4 & 0.3 & 0.17 & 1 \\
  18757 & -0.28 & 0.32 & 0.25 & 11.5 &  68.6 & -79.2 &-27.7 &  93.3 &  9.1 & 0.3 & 0.36 & 2 \\
  19445 & -1.99 & 0.47 &      & 14.0 &-157.2 &-122.0 &-67.1 & 210.8 & 12.4 & 1.6 & 0.67 & 3 \\
  22879 & -0.86 & 0.44 & 0.42 & 14.0 & 108.9 & -86.0 &-44.8 & 130.7 & 10.1 & 0.6 & 0.45 & 2 \\
  25329 & -1.84 & 0.42 & 0.24 & 13.0 &  40.2 &-189.0 & 19.9 & 183.9 &  8.6 & 4.4 & 0.85 & 3 \\
  29907 & -1.55 & 0.29 & 0.63 & 14.0 & 381.2 &-142.0 & 27.8 & 395.7 & 44.8 & 3.0 & 0.94 & 4 \\
  30649 & -0.47 & 0.35 & 0.32 & 12.0 &  58.5 & -80.6 & -9.6 &  85.7 &  8.9 & 0.0 & 0.35 & 2 \\
  30743 & -0.45 & 0.14 & 0.13 &  5.0 & -25.8 &  -5.4 &-23.6 &  40.1 &  9.9 & 0.2 & 0.12 & 1 \\
  31128 & -1.49 & 0.34 & 0.44 & 13.0 &  59.4 & -97.1 &-26.1 & 102.1 &  8.9 & 0.3 & 0.42 & 2 \\
  34328 & -1.61 & 0.38 & 0.27 &  8.0 & 207.7 &-352.0 & 95.6 & 407.9 & 15.4 & 4.7 & 0.67 & 4 \\
  37124 & -0.44 & 0.32 & 0.30 & 14.5 & -28.7 & -46.5 &-43.7 &  65.2 &  9.0 & 0.6 & 0.20 & 2 \\
  43042 &  0.04 & 0.00 &-0.02 &  1.4 &  32.5 & -18.4 &-16.6 &  26.2 &  8.9 & 0.1 & 0.08 & 1 \\
  45282 & -1.52 & 0.37 & 0.60 & 14.0 & 245.8 &-186.0 &-43.7 & 297.0 & 15.9 & 8.2 & 0.91 & 4 \\
  52711 & -0.16 & 0.04 & 0.05 &  7.0 &  18.5 & -77.5 & -9.2 &  68.1 &  8.5 & 0.0 & 0.31 & 1 \\
  55575 & -0.36 & 0.17 & 0.20 &  8.5 &  79.5 &  -1.8 & 32.1 &  79.7 & 11.5 & 0.7 & 0.23 & 1 \\
  58855 & -0.32 & 0.12 &      &  4.5 & -25.4 & -15.1 & -4.2 &  35.8 &  9.4 & 0.0 & 0.12 & 1 \\
  59392 & -1.59 & 0.27 & 0.68 & 13.0 &-123.0 &-325.0 &-32.3 & 343.5 & 10.6 & 0.4 & 0.67 & 4 \\
  61421 & -0.01 & 0.06 & 0.01 &  2.0 &  -5.4 &  -8.3 &-18.8 &  20.1 &  9.0 & 0.2 & 0.05 & 1 \\
  62301 & -0.69 & 0.30 & 0.36 & 12.0 &   5.5 &-108.0 &-23.4 & 100.1 &  8.5 & 0.2 & 0.45 & 2 \\
  64606 & -0.89 & 0.37 & 0.47 & 12.5 &  82.4 & -65.2 &  1.4 &  91.3 &  9.5 & 0.1 & 0.33 & 2 \\
  65583 & -0.73 & 0.39 & 0.46 & 12.5 &  12.0 & -88.6 &-30.9 &  82.5 &  8.5 & 0.3 & 0.36 & 2 \\
  67228 &  0.12 & 0.00 &-0.13 &  4.7 & -32.8 &   4.1 &-15.9 &  46.1 & 10.7 & 0.1 & 0.15 & 1 \\
  68017 & -0.40 & 0.34 & 0.28 & 13.0 &  48.1 & -60.4 &-39.6 &  71.5 &  8.8 & 0.5 & 0.25 & 2 \\
  69611 & -0.60 & 0.43 & 0.36 & 13.5 &  38.3 &-144.0 &-43.4 & 142.4 &  8.6 & 0.6 & 0.62 & 2 \\
  74000 & -2.00 & 0.35 & 0.16 & 11.0 &-245.7 &-362.0 & 59.5 & 440.3 & 21.1 & 2.7 & 0.77 & 4 \\
  84937 & -2.07 & 0.36 &      & 13.0 &-224.5 &-238.0 & -8.3 & 327.7 & 15.3 &10.1 & 0.99 & 4 \\
  90508 & -0.33 & 0.22 & 0.26 & 10.5 & -21.0 & -94.2 & 22.4 &  94.1 &  8.7 & 0.4 & 0.39 & 2 \\
  97320 & -1.18 & 0.36 & 0.40 & 13.0 & -73.1 & -20.0 &-37.5 &  89.4 & 11.0 & 0.5 & 0.27 & 2 \\
  99383 & -1.54 & 0.37 & 0.40 & 14.0 &  23.2 &-204.0 &129.0 & 237.1 &  8.9 & 7.3 & 0.79 & 3 \\
 102158 & -0.46 & 0.40 & 0.34 & 13.5 & 114.4 &-118.0 & 10.1 & 151.5 &  9.9 & 0.2 & 0.57 & 2 \\
 102200 & -1.24 & 0.30 & 0.57 & 14.0 & -86.5 &-131.0 &  6.4 & 155.5 &  9.5 & 0.2 & 0.61 & 3 \\
 103095 & -1.35 & 0.28 & 0.55 & 14.0 &-278.0 &-160.0 &-13.4 & 325.3 & 21.8 & 0.2 & 0.90 & 4 \\
 109358 & -0.21 & 0.07 &      &  7.5 &  30.8 &  -3.3 &  1.5 &  23.1 &  9.4 & 0.1 & 0.07 & 1 \\
 112758 & -0.43 & 0.36 & 0.28 & 12.5 &  75.7 & -34.1 & 16.4 &  73.5 &  9.8 & 0.3 & 0.23 & 2 \\
 114710 & -0.03 & 0.01 & 0.01 &  3.5 &  50.6 &   8.7 &  8.5 &  47.0 & 10.9 & 0.2 & 0.16 & 1 \\
 114762 & -0.71 & 0.33 &      & 12.5 &  82.0 & -70.5 & 57.8 & 113.6 &  9.5 & 1.3 & 0.34 & 2 \\
 117176 & -0.11 & 0.08 & 0.04 &  8.1 & -13.7 & -51.7 & -3.8 &  48.0 &  8.6 & 0.0 & 0.20 & 1 \\
\hline
\end{tabular}
\end{table*}  
 
\begin{table}
\tabcolsep0.9mm
\begin{tabular}{ccccccccccccc}
\hline
HD/BD & [Fe/H] & [Mg/Fe] & [Eu/Fe]& t, & U,  & V, & W, & $V_{pec}$, &
Ra, & $Z_{max}$, & e & êîä \\
      &  dex   &  dex    &  dex   &  Gyr    & km/s&km/s&km/s& km/s       & 
kpc & kpc      &   &     \\
\hline
 121560 & -0.43 & 0.15 & 0.14 &  5.0 &  29.2 & -20.1 & -3.1 &  21.9 &  8.8 & 0.0 & 0.08 & 1 \\
 122196 & -1.71 & 0.16 & 0.24 & 12.0 & 171.8 &-143.0 & 14.0 & 210.8 & 11.5 & 0.3 & 0.73 & 3 \\
 126053 & -0.35 & 0.14 & 0.14 &  9.0 & -22.4 & -15.2 &-39.3 &  46.7 &  9.3 & 0.5 & 0.11 & 1 \\
 130322 &  0.04 & 0.02 &-0.04 &  1.0 &   9.4 & -26.0 &-11.0 &  16.8 &  8.5 & 0.1 & 0.07 & 1 \\
 132142 & -0.39 & 0.31 & 0.26 & 12.5 & 108.4 & -55.9 & 19.6 & 111.6 & 10.5 & 0.4 & 0.37 & 2 \\
 134987 &  0.25 & 0.00 &-0.17 &  6.0 &  20.5 & -40.2 & 20.8 &  41.7 &  8.5 & 0.3 & 0.14 & 1 \\
 142373 & -0.57 & 0.22 & 0.23 &  8.5 &  41.8 &  10.4 &-68.2 &  72.8 & 11.3 & 1.3 & 0.16 & 1 \\
 144579 & -0.69 & 0.38 & 0.46 & 12.5 &  35.5 & -58.2 &-18.0 &  55.8 &  8.6 & 0.1 & 0.23 & 2 \\
 148816 & -0.78 & 0.41 & 0.30 & 12.0 & -86.5 &-262.0 &-79.2 & 280.4 &  9.7 & 5.4 & 0.89 & 4 \\
 157214 & -0.34 & 0.38 & 0.34 & 12.5 & -26.1 & -80.4 &-63.8 &  97.9 &  8.7 & 1.1 & 0.32 & 2 \\
 168009 & -0.03 & 0.04 &-0.08 &  9.0 &   4.4 & -61.7 &-22.5 &  54.5 &  8.5 & 0.2 & 0.24 & 1 \\
 176377 & -0.27 & 0.07 &      &  2.5 &  38.5 & -23.7 & -4.7 &  31.7 &  8.9 & 0.0 & 0.11 & 1 \\
 179957 & -0.01 & 0.05 & 0.03 &  8.0 &  67.2 & -44.0 & 35.5 &  78.4 &  9.4 & 0.7 & 0.23 & 1 \\
\hline
\end{tabular}
\end{table}  

\end{document}